\documentclass[letterpaper, 10 pt, conference]{ieeeconf}  

\IEEEoverridecommandlockouts                              

\usepackage{graphicx} 
\usepackage{amsmath} 
\usepackage{amssymb}  
\usepackage{multirow}
\usepackage{hyperref}

\usepackage{etoolbox}
\makeatletter
\patchcmd{\@makecaption}
  {\scshape}
  {}
  {}
  {}
\makeatletter
\patchcmd{\@makecaption}
  {\\}
  {.\ }
  {}
  {}
\makeatother

\graphicspath{ {./img/} }

\title{\LARGE \bf
Probabilistic Rainfall Estimation from Automotive Lidar
}

\author{Robin Karlsson$^{1 * \dagger}$, David Robert Wong$^{2}$, Kazunari Kawabata$^{2}$, Simon Thompson$^{2}$, and Naoki Sakai$^{3}$
\thanks{$^{1}$Graduate School of Informatics, Nagoya University, Aichi, Japan.}%
\thanks{$^{*}$Corresponding author: karlsson.robin@g.sp.m.is.nagoya-u.ac.jp}%
\thanks{$^{\dagger}$This work was done while employed at Tier IV}
\thanks{$^{2}$Tier IV Inc., Tokyo, Japan.}%
\thanks{$^{3}$The National Research Institute for Earth Science and Disaster Resilience, Ibaraki, Japan}%
\thanks{Code: \url{github.com/tier4/rainfall_modeling_open}
}
}

\begin{document}

\maketitle
\thispagestyle{empty}
\pagestyle{empty}

\begin{abstract}

Robust sensing and perception in adverse weather conditions remain one of the biggest challenges for realizing reliable autonomous vehicle mobility services. Prior work has established that rainfall rate is a useful measure for the adversity of atmospheric weather conditions. This work presents a probabilistic hierarchical Bayesian model that infers rainfall rate from automotive lidar point cloud sequences with high accuracy and reliability. The model is a hierarchical mixture of experts model, or a probabilistic decision tree, with gating and expert nodes consisting of variational logistic and linear regression models. Experimental data used to train and evaluate the model is collected in a large-scale rainfall experiment facility from both stationary and moving vehicle platforms. The results show prediction accuracy comparable to the measurement resolution of a disdrometer, and the soundness and usefulness of the uncertainty estimation. The model achieves RMSE 2.42\,mm/h after filtering out uncertain predictions. The error is comparable to the mean rainfall rate change of 3.5\,mm/h between measurements. Model parameter studies show how predictive performance changes with tree depth, sampling duration, and crop box dimension. A second experiment demonstrates the predictability of higher rainfall above 300\,mm/h using a different lidar sensor, demonstrating sensor independence.

\end{abstract}

\section{INTRODUCTION}
\label{sec:introduction}

Autonomous vehicles (AVs) hold the promise to revolutionize personal mobility and free up commuting time. However, to reap the full benefits expected by the public, and to realize a reliable and economically viable mobility service, the AVs need to be able to robustly operate autonomously in a wide range of uncontrollable environments and traffic situations. A part of this problem is sensing and perception in adverse weather, which remains one of the biggest challenge needed to be overcome in order to create an economically viable reliable system the public can rely upon.

Rainfall rate and meteorological optical range (MOR) are considered measurable components of atmospheric weather condition adversity, as stated in prior works for road scene definition~\cite{ulbrich15}, AV safety frameworks~\cite{koopman2018, guo2018, gyllenhammar20}, and lidar performance modeling~\cite{rasshofer2011, goodin2019}. While prior work has established the usefulness of knowing precipitation values, no previous work exists which presents a practical solution with sufficient accuracy for real-time estimation of rainfall rate from automotive lidar data.

This work demonstrates the use of data-driven probabilistic modeling to infer reliable values for rainfall rate based on temporal automotive lidar data from both stationary and moving vehicle platforms, achieving higher accuracy and granularity than prior state-of-the-art approaches \cite{heinzler19, lewandowski08, perin17}.  All code and models to reproduce the presented findings will be made available upon publication.

Our contributions are twofold:
\begin{itemize}
 \item A probabilistic hierarchical mixture of experts model achieving state-of-the-art accuracy and sound uncertainty estimation for rainfall rate prediction from automotive lidar data in real-time, across the range of naturally occurring rain 0 to 300\,mm/h.
 \item An investigation of model parameter influence on predictive performance based on experimental data for rainfall rates between 0\,mm/h and 60\,mm/h.
\end{itemize}

\begin{figure}
  \centering
  \includegraphics[width=0.45\textwidth]{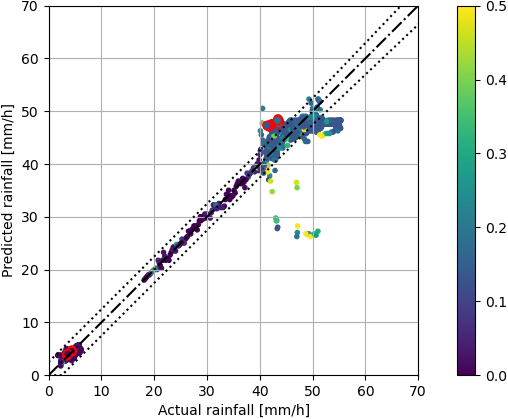}
  \caption{Predicted rainfall results represented by dots. Color represents the uncertainty as error probability, namely the likelihood that the actual rainfall value is not within $\pm$5\% of the predicted rainfall value. The dotted lines indicate error bounds for the lower of $\pm$2.5\,mm/h or $\pm$5\% error. RMSE is 2.42\,mm/h after excluding uncertain predictions above 25 \% error probability (95.5 \% points remaining). Validation samples have red edges.}
  \label{fig:plot_all_crop_10_tree_2_sampling_150}
\end{figure}

\section{RELATED WORK}
\label{fig:related_work}

\subsection{Precipitation estimation using lidar}

Prior meteorological works \cite{rensch70, shipley74, chimelis82, stow91} have derived theoretical rainfall models based on physical principles, involving optical scattering and estimates of droplet shape distributions~\cite{marshall48}. These rainfall rate models are experimentally verified based on attenuation coefficient measurements obtained from long-range and high-power meteorological lidar measurement data.

A more recent work by Lewandowski~et~al.~\cite{lewandowski08} models the effect of precipitation as increased attenuation of lidar beam energy due to absorption, random scattering from water droplets~\cite{hassler98, deirmendjian69}, and decreased reflectivity from wet surfaces~\cite{lekner88, measures84}. Parameters of the model are obtained by fitting ground truth rainfall measurements from an optical disdrometer with a priori known relations for power intensity and range measurements~\cite{klett81}. Experimental results show a relative error of 55\,\% between lidar and disdrometer values across the measurement range of 0.1-100\,mm/h. Our proposed data-driven model significantly improves accuracy while also being applicable with automotive lidar.

A work by Perin~\cite{perin17} using low-cost automotive lidar demonstrates that it is possible to model rainfall rates above 10\,mm/h based on beam intensity and range estimation using linear and power law models fitted to stationary automotive lidar experimental data. Due to the simple modeling approach, the results show high variability and hence low accuracy when compared to our proposed model.

\subsection{Weather detection using lidar}

Heinzler~et~al.~\cite{heinzler19} presents a supervised learning method to train a support vector machine (SVM) model that classifies the weather condition as `clear', `fog', or `rain'. Their training data consists of feature vectors computed from temporal and spatial lidar point cloud statistics, with target labels being from the three classes. The model is trained and evaluated on climate chamber and road experiment data. Our proposed model extends their work, by enabling quantitative estimation of the severity of adverse weather conditions, such as rainfall rate. Dannheim~et~al.~\cite{dannheim14} also presents a method using both lidar and camera to detect rain, snow, and fog. Other works attempt to quantify lidar degradation without explicitly quantifying the rainfall rate. For example, Hasirlioglu~et~al.~\cite{hasirlioglu16} presents a physics-based model, while Zhang~et~al.~\cite{zhang21} presents a learning-based method.

\subsection{Precipitation lidar modeling}

Several recent works \cite{rasshofer2011, goodin2019, li2020} have proposed lidar models based on physical principles that estimate the degradation of automotive lidar sensor measurements given an estimate of precipitation and/or fog is already known. The models are verified to be accurate from comparison with lidar data presented by Filgueira~et~al.~\cite{filgueira16}. A method to quantify rainfall rate from automotive lidar data as presented by us is thus valuable, as such a method would indirectly allow accurate estimation of lidar performance in adverse weather conditions and therefore increase the robustness of the AV.

\section{RAINFALL MODELING}
\label{sec:probabilistic_precipitation_model}

In this section, we present a data-driven probabilistic hierarchical machine learning model that is trained to predict rainfall rate and uncertainty in the prediction. The dataset consists of feature vectors generated from temporal sequences of lidar point clouds, and ground truth rainfall rate measured by a disdrometer.

Prior work in meteorology~\cite{lewandowski08} has pursued theoretical models for inferring rainfall rate using high-power lidars and strong a priori assumptions on sensing conditions. In practice, measurements from automotive lidar are relatively noisy due to their low-power eye safety requirement, and assumptions such as a homogeneous atmosphere and undisturbed beam propagation are broken, implying that models based on the physical sensing process are ill-suited for the automotive domain~\cite{perin17}.

For this reason, we pursue a data-driven machine learning approach to overcome the practical limitations entailed by physics-based models and learn models without limiting modeling assumptions. Our data-driven approach follows prior work made by Heinzler~et~al.~\cite{heinzler19} for weather detection, and Li~et~al.~\cite{li2020} for lidar performance estimation in fog.

\subsection{Lidar rainfall modeling characteristics}

\begin{figure}
  \centering
  \includegraphics[width=0.48\textwidth]{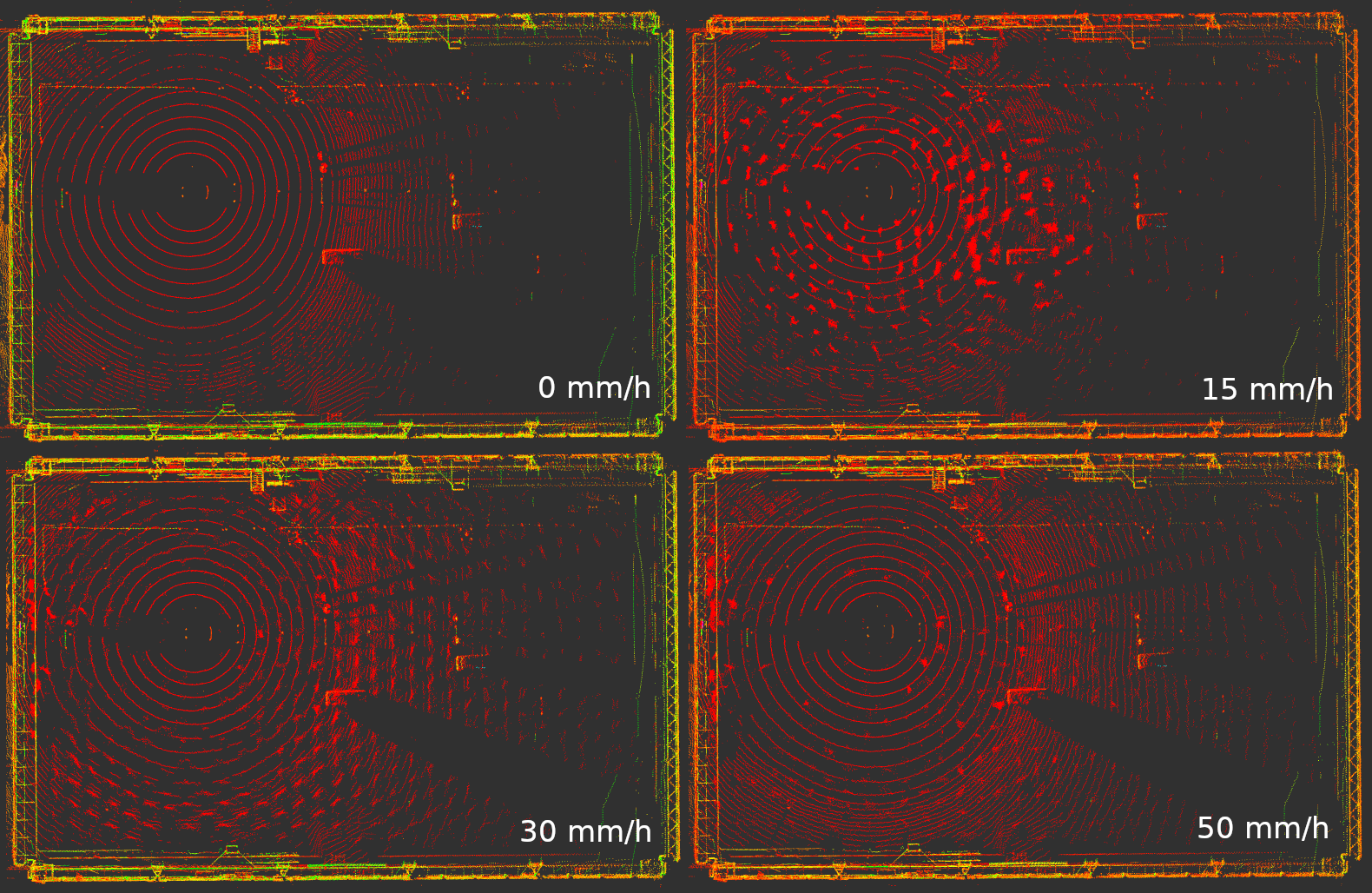}
  \caption{Visualization of point cloud scans for different rainfall rates from the top-down perspective. Color represents intensity scaled between the minimum (red) and maximum (purple) values within a scan. The noise pattern changes nonlinearly with increasing rainfall rate.}
  \label{fig:topdown_lidar}
\end{figure}

The effect of precipitation on a point cloud scan is directly observable in terms of reduced intensity of laser returns, reduced maximum sensing range, increased range estimation noise, and false positives due to raindrop scattering, The combined effect of these factors results in a nonlinear change in the overall spatial and intensity distribution of points \cite{rasshofer2011, filgueira16}, and is the motivation for introducing hierarchical modeling in order to train different models for different rainfall ranges. Fig.~\ref{fig:topdown_lidar} visualizes how the point cloud noise pattern changes with rainfall rate. A higher rainfall rate does not decidedly result in more noise, but rather a different kind of noise pattern related to varying droplet size and interactions.

\subsection{Quantification of uncertainty}

Uncertainty quantification is fundamental for safety-critical functionality and necessary for risk-aware decision making within subsequent systems, allowing estimation of the value of information in regard to its uncertainty. The estimated uncertainty distribution can be used to measure error probability or the likelihood that a predicted value is within an error margin of the true value. The error probability value is used in an outlier filter to improve predictive robustness by removing unmodelable inputs.

While recent work \cite{sensoy18} argues that discriminative models like neural networks can learn to output prediction uncertainty for categorical Dirichlet distribution outputs including an explicit `unknown' class, the approach cannot be applied to regression problems without discretizing the value range.

In this work, we estimate prediction uncertainty for continuous values using a Bayesian modeling approach. In Bayesian models, the weights or parameters themselves are assumed uncertain and thus modeled as probabilistic distributions. The uncertainty of a prediction is measured as the variance of the resulting output generated using all possible parameters and weighted by model probability.

The model parameters can be estimated using a deterministic approximation method called variational inference \cite{jordan99, blei17} that estimates a lower bound approximation of the unknown true probability distributions through optimizing parametric distributions such as the Gaussian distribution. The probabilistic model output is computed by integrating over all model parameters represented by the optimized distributions. 

\subsection{Probabilistic hierarchical rainfall model}

\begin{figure}
  \centering
  \includegraphics[width=0.45\textwidth]{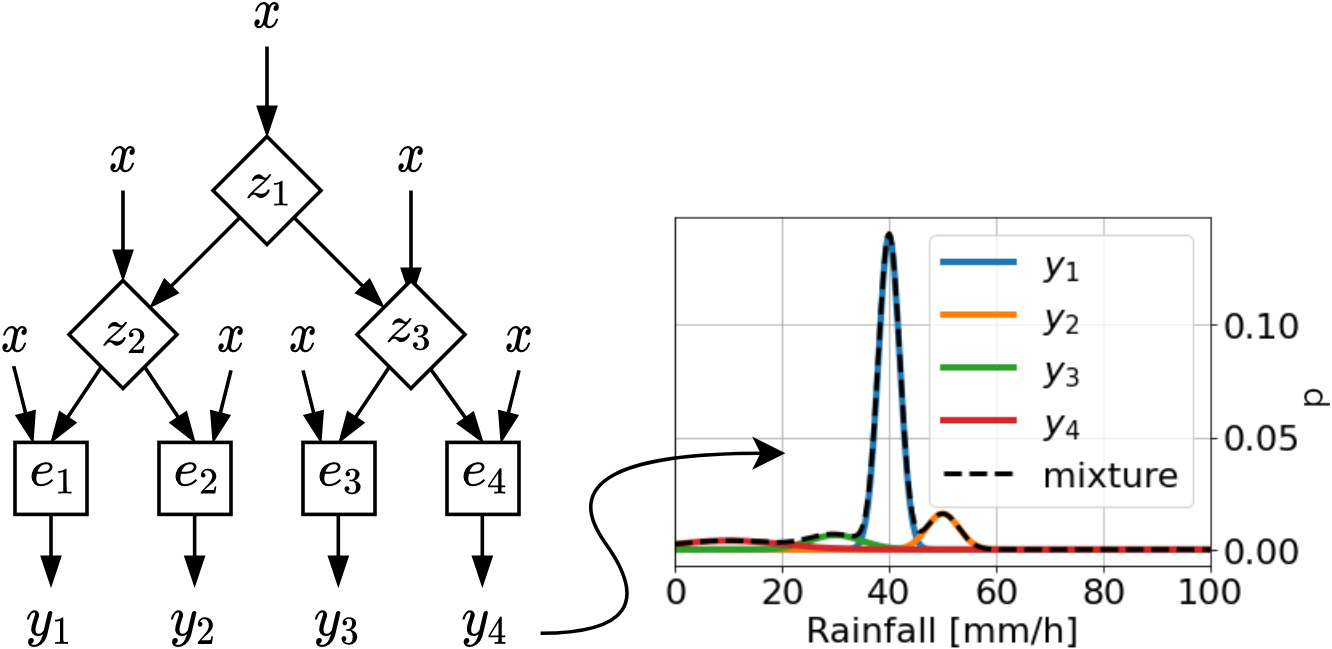}
  \caption{A hierarchical mixture of experts model of depth two, consisting of three gating functions $z_1 \dots z_3$ and four expert models $e_1 \dots e_4$. Each gating function $z_k$ computes a binary probability of sample $x$ belonging to the left or right branch of the tree. Propagating all probabilities through the tree gives the probability $P_m$ of sample $x$ belonging to each of the expert models $e_m$. Each expert model $e_m$ outputs a prediction $y_m = \mathcal{N}(\mu_m, \sigma^2_m)$, representing a Gaussian distribution of the predicted rainfall rate. The final model output is represented by a mixture of Gaussians probability distribution~Eq.~(\ref{eq:mixture_of_gaussians}).}
  \label{fig:model_diagram}
\end{figure}

One family of models that satisfy the requirements for precipitation modeling is the mixture of experts model~\cite{jacobs91}, also known as a probabilistic decision tree~\cite{bishop06}. Fig.~\ref{fig:model_diagram} provides a diagram illustrating this family of models. The tree nodes are composed of binary decision gating nodes. Each gate node $z_k$ predicts the probability $P_k$ that the target label $y$ of a feature vector $x$ is higher than the threshold value $h_k$ of the gate. In other words, the gating node learns the probability of $x$ branching left or right. Each tree leaf consists of a expert node $e_m$ which outputs a distribution $\mathcal{N}(\mu_m, \sigma^2_m)$ for feature vector $x$, representing the likelihood of predicting $y$.

Propagating the probabilities $P_1, \dots, P_K$ down the tree, results in probability values $P_1, \dots, P_M$, denoting the probability $P_m$ that $x$ belongs to the target value range expert node $e_m$ is trained to predict. The model prediction is computed by summing the predicted distributions of all experts in accordance with the probability the gating nodes have assigned of $x$ belonging to the target value range of expert $e_m$:

\begin{equation}
 \label{eq:mixture_of_gaussians}
 p(y) = \sum^M_{m = 1} P_m \: \mathcal{N}(y | \mu_m, \sigma^2_m).
\end{equation}

The gate nodes are modeled as variational logistic regression models~\cite{jaakkola00}, which learn a multivariate Gaussian distribution over model parameters, by optimizing an analytically integrable variational lower bound approximation of the posterior distribution to maximize the likelihood of the data. An essential point is that representing model parameters as integrable distributions (i.e. not point estimates) allows integration over all parameter values, meaning the model output represents the combination of all possible models in proportion to the likelihood of parameter values.

If for a feature vector $x$, transformed by a basis function $\phi(x)$, the model parameter distributions result in a highly varying result, the output value is uncertain. On the other hand, if $x$ results in a stationary result, the output value is more likely to be correct, and thus more certain. For both cases, this uncertainty is represented by an explicit probability value. Integration over model parameter distributions, optimized according to data $\mathcal{D}$, results in the following analytical function for binary classification probability:

\begin{equation}
 p(z=True|x, \mathcal{D}) = \sigma \left( \kappa(\sigma^2_a) \mu^T_N \phi(x) \right )
\end{equation}

\noindent where $\kappa(\sigma^2_a) = \left ( 1 + \pi \sigma^2_a / 8 \right )^{-1/2}$, $\sigma^2_a = \phi(x)^T \Sigma_N \phi(x)$, $\sigma()$ is the sigmoid function, and $\mu_N$ and $\Sigma_N$ represent the mean vector and covariance matrix of the parameter distributions.

The expert nodes are modeled as variational linear regression models~\cite{bishop06}, which similarly to the variational logistic regression gate nodes, optimize analytically integrable lower bound approximations for model parameter distributions. After integrating model parameters, the model output becomes a Gaussian distribution for the predicted value $\hat{y}$:

\begin{equation}
 p(y|x, \mathcal{D}) = \mathcal{N}(y|\mu^T_N \phi(x), \sigma^2(x))
\end{equation}

\noindent with estimated variance or uncertainty of $\hat{y}$ computed as

\begin{equation}
 \sigma^2(x) = \beta^{-1} + \phi(x)^T \Sigma_N \phi(x)
\end{equation}

\noindent where $\beta$ is an optimized noise precision parameter.

Different from a conventional discriminative decision tree, which only does hard expert assignments resulting in a single expert prediction, the probabilistic gating tree allows the expression of expert selection uncertainty, and the probabilistic expert nodes explicitly model the uncertainty in the target value. The probabilistic output allows a better distinction between reliable predictions and uncertain predictions better be ignored. Refinement of results based on uncertainty information is demonstrated in Sec.~\ref{sec:experiments}.

\subsection{Model learning}
\label{sec:model_learning}
Prior to learning, the expected rainfall rate range is partitioned into $M$ subdomains, that is, one subdomain per expert node $e_m$. Similarly, each gate node is associated with a rainfall rate threshold value $h_k$. The threshold value of lowest-level gate nodes ($z_2$ and $z_3$ in Fig.~\ref{fig:model_diagram}) is set to coincide with border values between expert ranges, effectively allowing the lowest-level gates to predict experts (threshold for $z_2$ corresponding to the upper range of $e_1$ and lower range of $e_2$ in Fig.~\ref{fig:model_diagram}). Using Fig.~\ref{fig:model_diagram} to illustrate a threshold and expert range assignment: $z_1 = 20$, $z_2 = 10$, $z_3 = 40$, $e_1 \in (0, 10)$, $e_2 \in (10, 20)$, $e_3 \in (20, 40)$, and $e_4 \in (40, \infty)$ 

The mixture of experts model is learned in two steps. First, the nodes $z_k$ in the gating tree are trained to correctly predict if the target value $y^{(i)}$ of feature vector $x^{(i)}$ is larger than the threshold value $h_k$ of the node. This training process is repeated for all gate nodes. Samples already assumed predicted by prior gate nodes are pruned to increase the discriminative ability of the model on relevant samples (using Fig.~\ref{fig:model_diagram} to illustrate; $z_2$ only learns from samples where $y^{(i)} < h_1$). Random duplication of samples is done to achieve class balance. The second step involves training expert models $e_m$ to output a prediction value distribution $\mathcal{N}(\mu_m, \sigma^2_m)$ maximizing the likelihood of the target values $y^{(i)}$. Expert models are only trained on samples $x^{(i)}$ with $y^{(i)}$ within the range associated with each expert model $e_m$.

Both gate and expert models are learned using variational inference \cite{jordan99, blei17}, which optimizes a lower bound multivariate Gaussian distribution as an approximation of the true model parameters which maximizes the likelihood of data.

A favorable property of Bayesian machine learning is that one can avoid the over-fitting problem and determine the optimal model complexity on training data alone~\cite{bishop06}. The reason is that Bayesian methods marginalize over all possible values of the model parameters in accordance with their likelihood, while conventional maximum likelihood estimation results in parameter point estimates. This allows models to be compared without a validation set. However, the available data may not encompass the entire range of plausible data, meaning a separate set of hold-out data is used to cross-validate the final model~\cite{bishop06}.

\subsection{Model inference}

Inference on a sample $x$ using a learned mixture of experts model is likewise performed in two steps. First, starting from the top of the tree, the probability of target value $y$ for $x$ being larger than the threshold of first gate node (i.e. above threshold $h_1$ of gate node $z_1$ in Fig.~\ref{fig:model_diagram}) is calculated, and similarly for the following gate nodes (i.e. $z_2$ and $z_3$ in Fig.~\ref{fig:model_diagram}). The computed gate probabilities $P_1, \dots, P_K$ are propagated down the tree, eventually reaching the expert node leaves, where the resulting probability $P_m$ denotes the probability of $x$ belonging to the expert model $e_m$. Using Fig.~\ref{fig:model_diagram} as an example, the probability of $x$ belonging to expert model $e_3$ equals $P_3 = P(z_1 = True) P(z_3 = False)$. Secondly, the output of all experts $\mathcal{N}(\mu_m, \sigma^2_m)$ is computed and combined into a mixture of Gaussian distribution $p(y)$ following~Eq.~(\ref{eq:mixture_of_gaussians}).

\subsection{Lidar point cloud feature generation}
\label{sec:lidar_point_cloud_feature_generation}

Feature vectors $x$ for sample $i$ are generated from a sequence of $J$ lidar point scans $P^{1:J} = \{P^1, \dots , P^J\}$. First, for each point cloud $P^j$,  all points associated with known static and dynamic objects are filtered out~\cite{zermas17}, leaving only true noise points. The remaining points within a crop box centered on the lidar are extracted. The following four features are then computed for each of the extracted point clouds: total number of points $N$, average intensity $\tilde{p}$, average radial distance from lidar device $\tilde{r}$, and normalized minimum spanning tree (MST) length $\tilde{l}$. Next, the mean and standard deviation is computed from the values of all features from all point clouds, resulting in the statistical feature vector $x^{(i)} = [\mu_{N}, \sigma_{N}, \mu_{\tilde{p}}, \sigma_{\tilde{p}}, \mu_{\tilde{r}}, \sigma_{\tilde{r}}, \mu_{\tilde{l}}, \sigma_{\tilde{l}} ]^T$.

The normalized MST length $\tilde{l}$ is a measure of the degree of clustering of the point cloud. The usefulness of normalized MST length is that it quantifies the character and degree of clustering within a point cloud; $\tilde{l} \approx 1$ indicates points being distributed uniformly, $\tilde{l} < 1$ indicates the existence of clustered points, while $\tilde{l} > 1$ indicate points being distributed towards the edges of the crop box. Computing the value for $\tilde{l}$ involves first computing the standard MST length between all extracted points, using for example Prim's algorithm. This length is divided by the corresponding MST length of the same number of uniformly distributed points.

The learned model is expected to be sensor-specific, as different lidars have different inherent noise characteristics.

\section{EXPERIMENTS}
\label{sec:experiments}

\subsection{Experiment facility and sensor setup}
\begin{figure}
  \centering
  \includegraphics[width=0.43\textwidth]{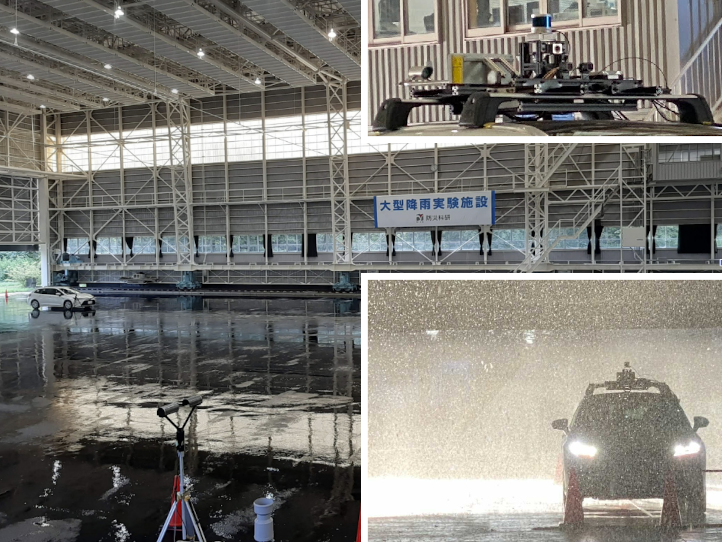} 
  \caption{The rainfall facility at NIED \cite{NIED} where lidar point cloud and rainfall measurement data was collected from a stationary and moving vehicle in 15\,mm/h and 50\,mm/h rainfall rate setup. The rainfall area encompasses the entire hall which measures 65.4~m$^2$. The disdrometer can be seen in the bottom-left of the figure. The sensor setup is shown in the top-right corner.}
  \label{fig:nied_exp_setup}
\end{figure}

The presented precipitation modeling approach is evaluated on real lidar and rainfall measurement data collected during a set of experiments in the large-scale rainfall simulator facility operated by The National Research Institute for Earth Science and Disaster Resilience (NIED) located in Tsukuba, Japan \cite{NIED}. The facility is capable of producing realistic rainfall with mean intensity between 15 and 300~mm/h. An image of the facility and experimental setup is shown in Fig.~\ref{fig:nied_exp_setup}.

Experiments were performed in two sets, each having a different sensor setup and tested rainfall rate range. The first set of experiments is considered the primary experiments, for which detailed methodology and results are presented and the experiment set is referred to unless explicitly stated otherwise. The second set of experiments applies the same methodology and only the final results are presented.

The first experiment set uses one Velodyne VLS-128 Alpha Prime lidar sensor mounted on a vehicle roof rack, recording data at 10 Hz. The lidar point clouds and rainfall measurement data are collected over four sessions. The first two sessions consist of data collected from a stationary vehicle in 15\,mm/h, 30\,mm/h, and 50\,mm/h rain for 10 minutes each. The final two sessions collect data from a moving vehicle in the same rainfall rates for 5 minutes each, allowing the rainfall rate to stabilize over 3 minutes before recording starts. The moving vehicle follows a figure-of-eight trajectory. We extract a 20 second segment from the middle of each recording session as validation data.

A stationary OTT Parsivel disdrometer is used to measure rainfall rate at 0.1\,Hz. The disdrometer is advertised to measure rainfall intensities ranging from 0.001\,mm/h to 1200\,mm/h with $\pm$5\,\% accuracy~\cite{parsivel}. However, it has been reported that similar laser precipitation monitors tend to overestimate rainfall intensities by 19.2\,\% to 37.2\,\%~\cite{lanzing06}, meaning care must be taken when comparing measurement as well as trained models.

The second experiment set uses a Velodyne Ultra Puck VLP-32C sensor. Additional data for 160\,mm/h and above 300\,mm/h rainfall rates are collected from a stationary and moving vehicle. For extreme rainfall rates, we note that the measured ground truth rainfall values are closer to 400\,mm/h than the expected 300\,mm/h range, as laser-based disdrometers tend to overestimate high rainfall rates~\cite{lanzing06}.

We collected the lidar data using the multiple echo sensor mode. Multi-echo registration allows registration of up to two laser returns per beam, ensuring that points caused by raindrop reflections will not be filtered out due to returns from stronger reflections by solid obstacles further ahead. Using multi-echo registration provides denser and more consistent raindrop point clouds, especially in the vicinity of solid obstacles.

\begin{figure}
  \centering
  \includegraphics[width=0.45\textwidth]{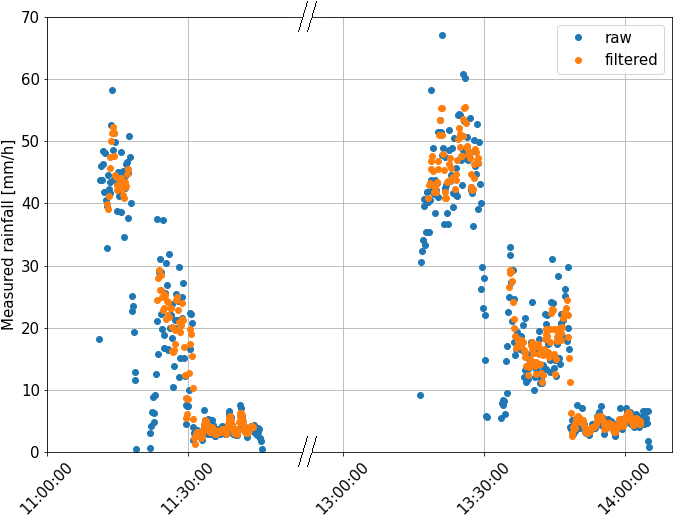}
  \caption{Rainfall measurement data collected by the Parsivel disdrometer. Blue points depict raw measurements. Orange points represent filtered measurements used as ground truth rainfall target values for modeling. The first and second sequences correspond to moving and stationary experiments.}
  \label{fig:rainfall_meas}
\end{figure}

\subsection{Data preprocessing}

The lidar data are preprocessed into noise and rainfall measurements as follows. First, we create a crop box centered on the lidar sensor and extract the points within the box. Secondly, all returns known to be caused by static and dynamic objects are removed using a map and point cloud semantic segmentation~\cite{hu2020}, leaving only noise points originating from rain droplet reflections. Thirdly, we associate a target rainfall value with the extracted point cloud by linearly interpolating between filtered ground truth rainfall measurements provided by the disdrometer.

Next, we generate independent training samples from a sequence of noise and rainfall measurements (i.e. 100 frames or 10 seconds). We compute statistical feature vectors from the sequential noise measurements as explained in Sec.~\ref{sec:lidar_point_cloud_feature_generation}, reducing the sequence into a feature vector $x$ and mean rainfall target value $y$. The set of all samples are represented by a dataset $\mathcal{D} = (X, Y)$, where feature vectors and rainfall target values are concatenated into a data matrix $X = \left [ x^{(1)}, \dots, x^{(N)}  \right]$ and target vector $Y = \left[y^{(1)} \dots y^{(N)}\right]$.

The raw and filtered rainfall rate measurements obtained from the Parsivel disdrometer are shown in Fig.~\ref{fig:rainfall_meas}. The raw measurements are filtered in order to reduce noise. First, a Savitzky-Golay filter of polynomial order 2 and window length 9 is applied to the sequence of raw measurement values. Secondly, 10 measurements (i.e. 100 sec) are cut from the beginning and end of each experiment segment in order to remove unstable rainfall measurements between rainfall rate adjustments. The final filtered values used as ground truth values $y$ exhibit a mean change of 3.5\,mm/h between every 10 sec measurement. This value is taken as the lower limit one can expect to model instantaneous rainfall rate given the temporal resolution of the Parsivel disdrometer.

\subsection{Model performance evaluation experiments}
\label{sec:model_evaluation_experiments}

\begin{figure*}[h!]
  \centering
  \includegraphics[width=1.\textwidth]{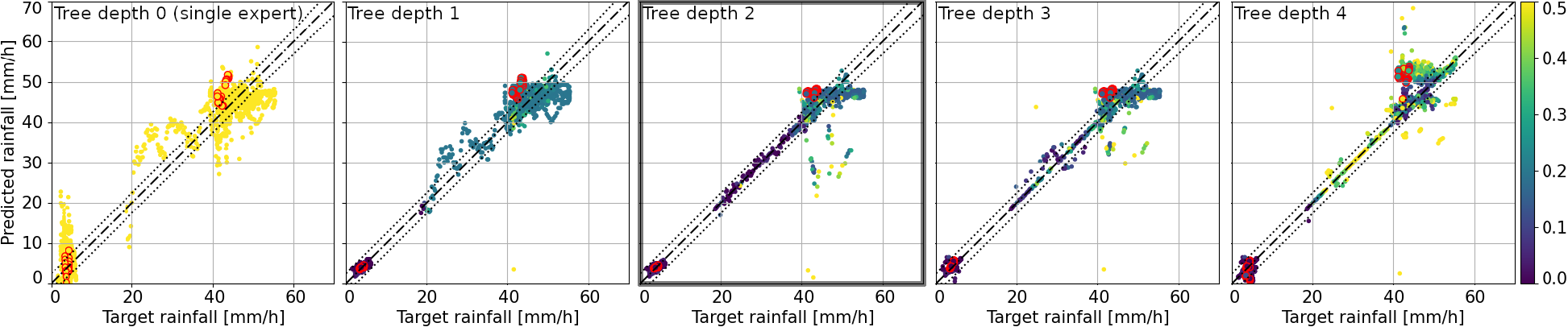}
  \caption{Summary of model tree depth experiment results. Increasing the gating tree depth is found to increase rainfall modelability as range-dependent nonlinear rainfall noise characteristics can be divided and learned by a larger set of expert models, in turn improving the overall predictive performance of the mixture of experts model. However, increasing the gating threshold resolution above the limit of modelability causes the gating tree to increasingly mispredict the best expert node for a sample, resulting in an increase of uncertain erroneous predictions. A tree depth of two (highlighted frame) is found to strike the best balance between prediction accuracy and uncertainty estimation (RMSE 2.53\,mm/h while retaining 93.1\,\% of samples). The leftmost figure demonstrates that a single model is insufficient for rainfall rate modeling, resulting in low predictive performance (RMSE 4.63\,mm/h) and indiscriminately high uncertainty (mean error prob. 83\%). This finding confirms the hypothesis that hierarchical modeling is an essential component for accurate and reliable rainfall modeling. Samples outlined in red are validation samples.}
  \label{fig:model_exp_summary}
\end{figure*}

\begin{figure*}[h!]
  \centering
  \includegraphics[width=0.80\textwidth]{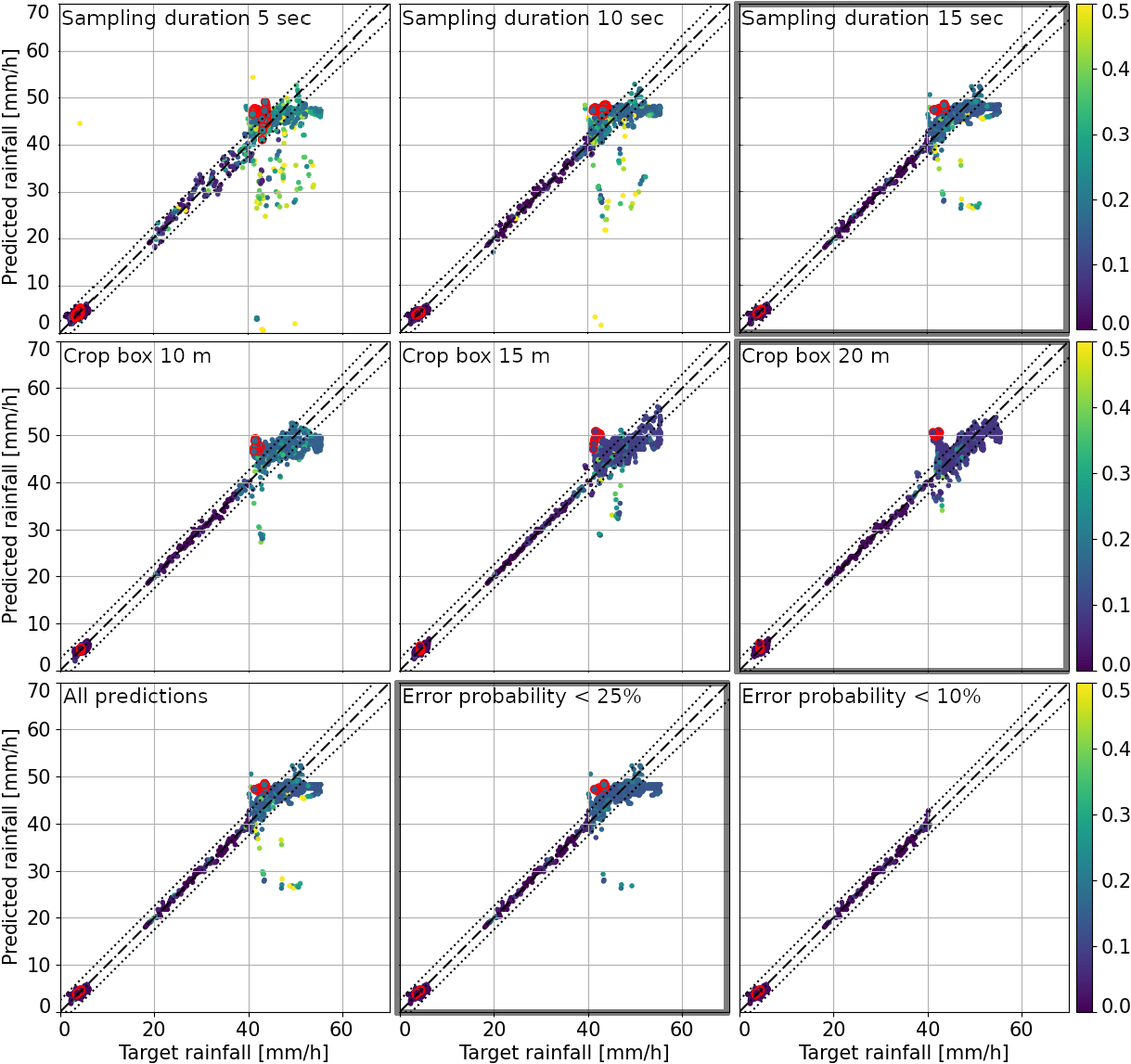}
  \caption{Summary of feature generation and uncertainty estimation experiment results. The first row shows that a longer sampling duration improves prediction accuracy and uncertainty estimation from RMSE 3.05\,mm/h (5 sec) to 2.42\,mm/h (15 sec) retaining 81.2\% and 95.5\,\% of points, respectively. Longer sampling duration is believed to significantly reduce the statistical variance of inferred features. The second row shows how increasing the crop box dimension improves prediction accuracy and uncertainty estimation from RMSE 2.37\,mm/h (10\,m) to 1.88\,mm/h (20\,m), retaining 96.4\% and 99.2\% of points, respectively. The number of rainfall-related points naturally increases with larger measurement volume, decreasing the statistical variance of features. The third row demonstrates how estimated uncertainty can be used to filter out a small number of uncertain predictions to improve overall accuracy. The leftmost plot shows all predictions of the baseline model configuration having RMSE 2.89\,mm/h. The following plots show the predictions remaining after filtering out uncertain ones, improving RMSE to 2.42\,mm/h and 0.656\,mm/h while retaining 95.5\.\% and 50.0\.\% of points, respectively. Highlighted frames represent the subjectively most favorable configurations as explained in~Sec.~\ref{sec:results}. Samples outlined in red are validation samples.}
  \label{fig:feature_exp_summary}
\end{figure*}

\begin{table*}
\centering
\caption{Experiment results. The first row denotes experiment, dataset, and static parameters. The following row specifies changing parameters. Results for predictive performance are given for all samples, as well as for predictions below 25\,\% and 10\,\% error probability (remaining samples in parentheses).}
\label{table:experiment_results}
\begin{tabular}{|r|r|l|l|l|l|l|}
\hline
\multicolumn{7}{|c|}{Tree depth experiments \(|\) All data, sampling duration: 10 sec, crop box: 10 m}                                                                                       \\ \hline
\multicolumn{1}{|l}{}                & \multicolumn{1}{l|}{} & Depth 0 (single expert) & Depth 1                   & \textbf{Depth 2}          & Depth 3          & Depth 4                  \\ \hline
\multirow{2}{*}{Metric: RMSE [mm/h]}                    & Train                 & 4.63                    & \textbf{2.86}             & 3.33                      & 2.91             & 4.01                     \\
                                     & Val.                  & 4.32                    & 4.27                      & \textbf{3.53}             & 3.57             & 6.41                     \\ \hline
\multirow{2}{*}{$<$25\% error prob.} & Train                 & \textendash \: (0\%)    & 2.75 (96.2\,\%)           & 2.53 (93.1\,\%)           & 2.45 (91.7\,\%)  & \textbf{2.10 (66.1\,\%)} \\
                                     & Val.                  & \textendash \: (0\%)    & 4.18 (95.5\,\%)           & \textbf{3.53 (100\,\%)}   & 3.57 (100\,\%)   & 5.80 (86.4\,\%)          \\ \hline
\multirow{2}{*}{$<$10\% error prob.} & Train                 & \textendash \: (0\%)    & \textbf{0.707 (44.9\,\%)} & 0.872 (50.0\,\%)          & 1.24 (45.3\,\%)  & 1.63 (51.4\,\%)          \\
\multicolumn{1}{|l|}{}               & Val.                  & \textendash \: (0\%)    & 0.28 (50.0\,\%)           & \textbf{0.265 (50.0\,\%)} & 0.911 (40.9\,\%) & 4.18 (54.5\,\%)          \\
\hline
\hline
\multicolumn{7}{|c|}{Sampling duration experiments \(|\) All data, tree depth: 2, crop box: 10 m}                                                   \\ \hline
\multicolumn{1}{|l}{}                & \multicolumn{1}{l|}{} & 5.0 sec                  & 10.0 sec                  & \textbf{15.0 sec}         & & \\ \hline
\multirow{2}{*}{Metric: RMSE [mm/h]}                    & Train                 & 4.41                     & 3.33                      & \textbf{2.89}             & & \\
                                     & Val.                  & \textbf{3.07}            & 3.53                      & 3.60                      & & \\ \hline
\multirow{2}{*}{$<$25\% error prob.} & Train                 & 3.05 (81.2\,\%)          & 2.53 (93.1\,\%)           & \textbf{2.42 (95.5\,\%)}  & & \\
                                     & Val.                  & \textbf{2.58 (86.0\,\%)} & 3.53 (100\,\%)            & 3.60 (100\,\%)            & & \\ \hline
\multirow{2}{*}{$<$10\% error prob.} & Train                 & 1.18 (47.5\,\%)          & 0.872 (50.0\,\%)          & \textbf{0.656 (50.0\,\%)} & & \\
\multicolumn{1}{|l|}{}               & Val.                  & 0.346 (45.3\,\%)         & \textbf{0.265 (50.0\,\%)} & 0.265 (50.0\,\%)          & & \\
\hline
\hline
\multicolumn{7}{|c|}{Crop box experiments \(|\) Stationary data only, tree depth: 2, sampling duration: 10 sec}                                                                                                          \\ \hline
\multicolumn{1}{|l}{}                & \multicolumn{1}{l|}{} & 10 m                      & 15 m            & \textbf{20 m}            & & \\ \hline
\multirow{2}{*}{Metric: RMSE [mm/h]}                    & Train                 & 2.55                      & 2.35            & \textbf{1.91}            & & \\
                                     & Val.                  & \textbf{4.24}             & 5.54            & 6.13                     & & \\ \hline
\multirow{2}{*}{$<$25\% error prob.} & Train                 & 2.37 (96.4\,\%)           & 2.12 (97.9\,\%) & \textbf{1.88 (99.2\,\%)} & & \\
                                     & Val.                  & \textbf{2.37 (100\,\%)}   & 5.54 (100\,\%)  & 6.12 (100\,\%)           & & \\ \hline
\multirow{2}{*}{$<$10\% error prob.} & Train                 & \textbf{0.600 (50.0\,\%)} & 1.78 (84.7\,\%) & 1.85 (87.7\,\%)          & & \\
                                     & Val.                  & \textbf{0.224 (50.0\,\%)} & 5.54 (100\,\%)  & 6.12 (100\,\%)           & & \\
\hline
\end{tabular}
\end{table*}

The following experiments evaluate how different model parameters impact predictive performance. Static model parameters for each experiment are specified in Table~\ref{table:experiment_results}.

\subsubsection{Effect of tree depth}

Five models with tree depths of one to four, as well as a single expert model representing tree depth zero, are trained and evaluated to evaluate the effect of tree depth on model performance. The tree of depth one consists of a single gate node $z_1$ and two expert nodes $e_1$ and $e_2$, having a single threshold value $h_1 = 20$\,mm/h. The tree of depth two has additional threshold values $h_{2,3} = (10, 40)$\,mm/h. The tree of depth three has further threshold values $h_{4,5,6,7} = (5, 15, 30, 60)$\,mm/h. The final tree of depth four greatly expands the previous tree with thresholds $h_{8,9,10,11,12,13,14,15} = (2.5, 7.5, 12.5, 17.5, 25, 35, 50, 70)$\,mm/h. See Fig.~\ref{fig:model_diagram} for a visualization of tree depth and threshold configuration. We space thresholds logarithmically to improve relative predictive performance for lower rainfall rates.

\subsubsection{Effect of sampling duration}

Four models are trained and evaluated on features computed from the same experimental data but of different frame sequences lengths 5\,sec, 10\,sec, and 15\,sec, in order to evaluate the effect of sampling duration on model performance.

\subsubsection{Effect of crop box size}

Three models are trained and evaluated on features computed from the same experimental data but of different crop box dimensions 10\,m, 15\,m, and 20\,m, in order to evaluate the effect of crop box dimension.

\subsubsection{Usefulness of estimated uncertainty}

We compare the change in predictive accuracy after removing predictions with error probability above 25 $\%$ and 10 $\%$. Prediction uncertainty is considered useful if removing a small subset of the most uncertain predictions improves overall accuracy.

\subsubsection{Secondary experiment set}

A model is trained and evaluated on data collected with another lidar sensor in higher rainfall rates. Sampling duration is 10\,sec. The crop box size is 10\,m. The model has a tree depth of three with thresholds $h_{1,2,3,4,5,6,7} = (112.7, 47.0, 197.0, 21.2, 77.5, 152.5, 246.2)$.

\section{RESULTS}
\label{sec:results}

\begin{figure}
  \centering
  \includegraphics[width=0.45\textwidth]{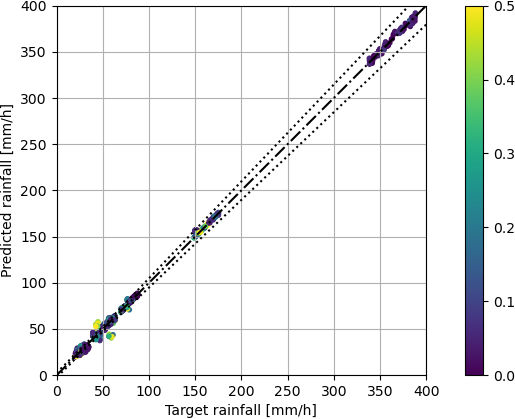}
  \caption{Rainfall predictions for the secondary experiment set. The result indicates good predictive performance also for higher rainfall rates, amounting to RMSE 2.02\,mm/h and 1.92\,mm/h with 88.5\,\% and 74.4\,\% point retention when filtering out samples having error probability above 25\% and 10\%, respectively.}
  \label{fig:plot_nied1}
\end{figure}

The experiment results are summarized in Table~\ref{table:experiment_results} and Fig.~\ref{fig:plot_all_crop_10_tree_2_sampling_150},\ref{fig:model_exp_summary}-\ref{fig:plot_nied1}. We find that the baseline model configuration with tree depth of two, and the longest tested sampling duration of 15\,sec, results in a good accuracy and point retention trade-off. In practice, we reckon this trade-off should be measured in terms of downstream task performance. Predictive performance after removing uncertain predictions is RMSE 2.42\,mm/h  (retaining 95.5\.\% of samples) for predictions with lower than 25\% error probability. The predictive accuracy is comparable to the disdrometer values which vary on average 3.5\,mm/h between 10 sec measurements. The secondary experiment results are shown in Fig.~\ref{fig:plot_nied1} confirm that a different lower-end 32 beam lidar sensor can also be used for rainfall rate modeling. The results also verify that high and extreme rainfall rates are also modelable.

Next, we summarize our findings regarding model parameters. A tree depth of two is found to strike a good balance between prediction accuracy and sample retention. A longer sample duration clearly improves predictive performance. However, the disdrometer measurements show that rainfall rate changes by 3.5\,mm/h on average every 10\,sec, implying that a distinction between instantaneous and mean rainfall rate modeling must be made. Increasing the crop box dimension also improves predictive performance. All experiments indicate that the estimated uncertainty can be used for improving predictive accuracy.

Code timing experiments over 1000 runs confirm that an inference rate of 1\,Hz is possible on a single CPU core. Converting moderate (524 points) and dense (2093 points) point clouds to feature vectors take on average 0.035\,sec and 0.43\,sec, respectively. Model inference takes 7.3\,msec.

\section{CONCLUSION}
\label{sec:conclusion}

This work presents a probabilistic hierarchical Bayesian model for predicting rainfall rate from a sequence of automotive lidar point clouds in real-time. The model is trained on data collected from a stationary and moving vehicle platform, using disdrometer ground truth rainfall rates and two different lidar sensors. A single model is demonstrating to cover the entire spectrum of naturally occurring rainfall upwards of 300\,mm/h. The most accurate model is verified to predict rainfall rates with RMSE  2.89\,mm/h, improved to 2.42\,mm/h when utilizing the estimated prediction uncertainty to filter out uncertain predictions, indicating accuracy comparable to the disdrometer measurement accuracy. Additional experiments establish relations between tree depth, sampling duration, crop box dimension, and predictive performance.

Future work includes optimizing the gating tree threshold values to match the actual rainfall noise pattern variations, and conducting experiments in natural rain and real road environments to demonstrate rainfall rate prediction in the presence of wind, interfering light sources, and objects.

\section{ACKNOWLEDGMENT}

We wish to express our gratitude to The National Research Institute for Earth Science and Disaster Resilience (NIED) for allowing us to use their facility for experiments.

\end{document}